\documentclass[a4paper,twoside]{article}
\newcommand{\affil}[1]{$^{\rm #1}$}
\usepackage[authoryear]{natbib}
\usepackage{amssymb, amsmath}
\bibpunct{(}{)}{;}{a}{}{,}
\usepackage{graphicx}
\date{} 
\title{\large\bf\flushleft The Adventures of the Rocketeer: \\ Accelerated Motion Under the
Influence of Expanding Space\footnotemark[1]}

\author{\parbox{\textwidth}{\flushleft
\vspace{-0.5cm}
{\it Juliana Kwan\affil{A,D}, Geraint F. Lewis\affil{A} and J. Berian James\affil{B,C}}\\
\vspace{0.4cm}
{\small \affil{A}\,Sydney Institute for Astronomy, School of Physics, A28, The
University of Sydney, NSW 2006, Australia}\\
{\small \affil{B}\,Institute for Astronomy, Royal Observatory, Edinburgh EH9 3HJ, UK}\\
{\small \affil{C}\,Dark Cosmology Centre, Niels Bohr Institute, University of
Copenhagen, Juliane Maries Vej 30, 2100 Copenhagen, Denmark} \\
{\small \affil{D}\,Email: jkwan@physics.usyd.edu.au}}}

\begin{document}
\twocolumn[
\begin{changemargin}{.8cm}{.5cm}
\begin{minipage}{.9\textwidth}
\vspace{-1cm}
\maketitle
%
%
\small{\bf Abstract:} It is well known that interstellar travel is
bounded by the finite speed of light, but on very large scales any
rocketeer would also need to consider the influence of cosmological
expansion on their journey. This paper examines accelerated journeys
within the framework of Friedmann- Lema\^{i}tre-Robertson-Walker
universes, illustrating how the duration of a fixed acceleration
sharply divides exploration over interstellar and intergalactic
distances.  Furthermore, we show how the universal expansion increases
the difficulty of intergalactic navigation, with small uncertainties
in cosmological parameters resulting in significantly large
deviations. This paper also shows that, contrary to simplistic
ideas, the motion of any rocketeer is indistinguishable from Newtonian
gravity if the acceleration is kept small.


\medskip
\medskip
\end{minipage}
\end{changemargin}
]
\small

\long\def\symbolfootnote[#1]#2{\begingroup%
  \def\thefootnote{\fnsymbol{footnote}}\footnotetext[#1]{#2}\endgroup} 

\section{Introduction} \symbolfootnote[1]{Research
  undertaken as part of the Commonwealth Cosmology Initiative (CCI:
  www.thecci.org), an international collaboration supported by the
  Australian Research Council}
The discovery of the accelerated expansion of the universe
\citep{Riess,Perlmutter} poses an interesting challenge to future
space travellers. In the currently favoured cosmological model, the
future of intergalactic travel appears grim as the universe becomes
dominated by dark energy and interesting structures, such as galaxies
and quasars, will have receded beyond the maximum distance that the
rocketeer can travel to, even after an infinite proper time. The `end
of cosmology' will be realised as the rocketeer infers the existence
of an `island universe', being able to reach fewer and fewer
destinations as these are accelerated beyond their event
horizon~\citep{Krauss,Loeb}.

In this paper, we consider the path of a rocketeer, who intends to
travel as far as possible in a Friedmann-Lema\^{i}tre-Robertson-Walker
(FLRW) universe with different cosmologies. The nature of the path
depends on several important features of the cosmology and this is
treated in terms of the limiting horizons alluded to by~\cite{Krauss,
Loeb} and also in use of phase portraits to describe the motion of the
rocketeer. We also address some of the problematic aspects of
travelling in an expanding universe, including the difficulty of
achieving a return trip due to the complicated dependence of the
motion of the rocketeer upon the evolution of the energy
densities. Furthermore, it is shown that although far more realistic
than constant acceleration, an initial burst of acceleration followed
by a `coasting' period is of little use if we wish to traverse
intergalactic distances. We also contribute to the current debate of
whether space \emph{really} expands (see~\cite{Lewis08,A08}
for recent developments) in the context of accelerated motion.

Several authors have explored unaccelerated paths through a FLRW
universe. See, for example,~\cite{Whiting,GE} for the general solution
for free particle motion in FLRW cosmologies as well
as~\cite{Barnes06} who clarifies the interpretation of these solutions
in the context of joining the Hubble flow. However, the treatment of
accelerated motion is uncommon, despite its potential for illuminating
the nature of expanding space as a physical phenomenon or a trick of
coordinates. Including the effect of acceleration increases the
complexity of the problem and indeed have only been discussed at
length in the context of cosmology by~\cite{Rindler60}
and~\cite{Heyl05}.~\cite{Rindler60} presented a novel method for
calculating accelerated paths by reducing the order of the
differential equations albeit by increasing the number of coupled
equations to solve. Of more relevence, ~\cite{Heyl05} considered the
problem of how far a rocketeer can travel in a human lifespan.

The layout of this paper is as follows; Section~\ref{bkgd} presents
the necessary background including the equations of motion of the
rocketeer, while Section~\ref{squiggly} addresses the return journey
of a rocketeer in a FLRW universe and its implications for the concept
of expanding space. Section~\ref{coasting} discusses the journey of an
intergalactic explorer who initially accelerates but then `coasts' at
a constant velocity close to the speed of light. Furthermore we also
discuss the importance of accelerated motion in the context of future
measurements in extragalactic astronomy. In the Appendix, we provide a
first order perturbative solution for the motion of the rocketeer.

\section{Background}\label{bkgd}
Accelerated motion in general relativity is most conveniently
approached by solving the geodesic equation with a four-acceleration
term. The equation of motion of the rocketeer is given by,  
given by,
\begin{equation} 
\epsilon^\mu \equiv \frac{Du^\mu}{d\tau} = \frac{du^\mu}{d\tau} + 
\Gamma^\mu_{\nu\sigma}\frac{dx^\nu}{d\tau}\frac{dx^\sigma}{d\tau}.
\label{ageo} 
\end{equation} 
where $\epsilon^\mu$ is the four acceleration, $u^\mu$ is the
four-velocity and $\tau$ is the proper time of the rocketeer. The
magnitude of the acceleration, $\epsilon$, corresponds to the force
experienced by the rocketeer. Equation~\ref{ageo} constitutes a set of
differential equations that are generally coupled for non-trivial
systems once an appropriate geometry has been inserted. Solving these
is the first step in determining the path of the rocketeer, but for
many cases this is analytically intractable and we have had to
recourse to numerical techniques extensively. We present a pertubative
solution in the Appendix, but this is unwieldy and cumbersome (whose
accuracy is limited to specific regimes of proper time). It is also
necessary to choose an appropriate means of slicing the time
coordinate for the FLRW model; in this case it is most expedient to
work in conformal coordinates. The path of the rocketeer asymptotes to
a null geodesic, as shown in~\citet{Heyl05}, which determines the
maximum comoving distance that can be travelled in an infinite proper
time. This is easily visualised when all photon paths trace out
straight lines oriented at $\pm$45$^\circ$, as in special relativity
and although this is not true in general for curved spacetimes, we can
enforce this behaviour through a conformal representation. Note that
such a transformation preserves angles at a point and as such paths
that are timelike in FLRW coordinates remain so in the conformal
geometry and vice versa. Since we are only considering motion in a
radial direction, we can express the FLRW metric
\begin{equation} 
ds^2 = a^2(\eta)( dr^2 - d\eta^2 ), 
\label{conf} 
\end{equation}
where $\eta$ is the conformal time and $a(\eta)$ is the scale factor
that governs the expansion of the universe. The conformal time is
simply a rescaling of the coordinate time (the time measured by a
stationary observer in a FLRW metric), via:
\begin{equation} 
a(t) \; d\eta \equiv dt. 
\label{eqn:conformal} 
\end{equation} 
Equation~\ref{conf} is only valid if we restrict our discussion to two
dimensions or universes without spatial curvature; if we wish to
consider a more general treatment, then it is necessary to perform a
transformation to fully conformal coordinates, in which the spatial
coordinates must be modified as well, (see~\cite{Infeld}
and~\cite{Lewis07} for a complete derivation).

Simply solving Equation~\ref{ageo} is not sufficient to derive the
motion of the rocketeer; she must also satisfy the following
constraints from the normalisation of $u^\mu$, the orthogonality of
$u^\mu$ and $\epsilon^\mu$ and the magnitude of $\epsilon^\mu$
(Equations \ref{frw_norm_expanded}, \ref{frw_acc_expanded} and
\ref{frw_perp_expanded}) that are inserted into their initial
conditions. These are expressed in the FLRW metric, as follows:
\begin{eqnarray}
g_{\mu\nu}u^\mu u^\nu = a(\eta)^2\left[{u^r}^2-{u^\eta}^2 \right] = -1,  \label{frw_norm_expanded} \\
g_{\mu\nu}\epsilon^\mu \epsilon^\nu = a(\eta)^2\left[{\epsilon^r}^2 -{\epsilon^t}^2\right]  = \epsilon^2, \label{frw_acc_expanded} \\
g_{\mu\nu} u^\mu \epsilon^\nu = u^r\epsilon^r - u^t\epsilon^t  =  0,   \label{frw_perp_expanded}
\end{eqnarray}
\noindent where the rocketeer is restricted to travel in the radial
direction only. Since this scenario is purely intended to address the
issue of inertial observers in expanding space, we have ignored the
details on how such a journey might be technically feasible and the
rocketeer is represented by a point particle of unspecified mass that
is capable of sustaining a constant acceleration for an indefinite
period.

\section{Accelerated motion}\label{squiggly}
To begin, we would like to consider the simplest case of accelerated
motion in an expanding universe: constant acceleration in a single
direction. For an universe described purely by special relativity, it
is trivial to show that the path of a rocketeer is a hyperbola of
the form:
\begin{equation}
t = (1/\epsilon) \sinh(\epsilon\tau); \: r = (1/\epsilon) \cosh(\epsilon\tau)
\label{eqn:milne}
\end{equation}
In this instance, constant acceleration means that the three-velocity
of the rocketeer rapidly approaches the speed of light and her path is
well approximated by that of a photon as \mbox{$\eta = r +
1/\epsilon$}, where $r$ is the comoving radial coordinate.

\begin{figure}[t] 
\begin{center}
\includegraphics[scale = 0.33,angle=270]{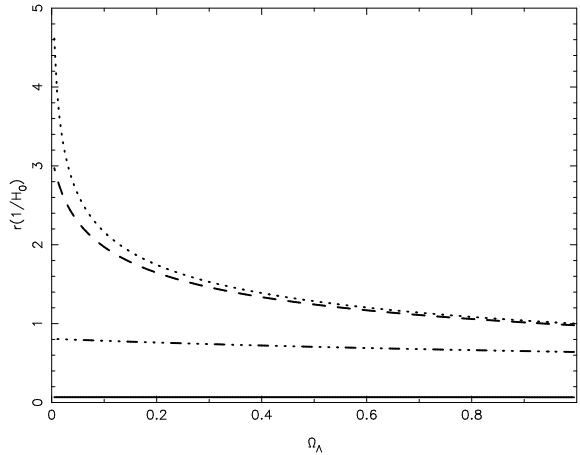}
\caption{Maximum radial comoving distances travelled by a rocketeer
with $\epsilon = g$ for proper times of 20 (solid), 23 (dot-dashed),
26 (dashed) and 29 (dotted) years. The rocketeer covers most of her
journey between $\approx$ 23--26 years as measured in her
frame. Indeed, after 26 years, the rocketeer hardly travels any
further in cosmologies with a non-zero $\Lambda$. From this point of
view, a non-zero cosmological constant is actually advantageous, since
for large values of $\Omega_\Lambda$, it takes very little time to
reach a substantial portion of the maximum distance accessible to an
intergalactic explorer. We have assumed $\Omega_\Lambda + \Omega_m =
1.0$ and that each journey starts from the current time in each
universe.
\label{fig:rocketeer}} 
\end{center}
\end{figure} 

A similar analysis may be performed in conformal coordinates to
determine how far the rocketeer can travel after an infinite time has
elasped because her path is bounded by a light cone projected from her
point of departure. For universes with a non-zero cosmological
constant, the conformal time in the future is finite after an infinite
cosmic time has elapsed~\footnote{In fact, any cosmology dominated by
a fluid with equation of state parameter less than -1/3 contains a
finite conformal future.}~\cite{Heyl05} showed that the furthest the
rocketeer can travel is a comoving distance equal to the remaining
conformal time for that cosmological model, which corresponds to
travelling for an infinite cosmic time; the transformation is purely
for convenience. As cosmic time approaches infinity, universes with a
non-zero cosmological constant, will have a scale factor that is well
approximated by an exponential (and is exact for $\Omega_{\Lambda,0} =
1 $ and $\Omega_{m,0} = 0 $). Thus, the conformal time remains finite
in all of these models, because the integral that defines conformal
time (Equation~\ref{eqn:conformal}) will always converge. However, the
rocketeer only reaches the maximum distance after an infinite proper
time has elapsed, but she is able to travel a comoving distance of
$\approx 4.63$ Gpc (99\% of the maximum conformal time, $\eta =
1.12339$) in $\approx$ 27 years in the currently favoured concordance
model with parameters $\Omega_{m,0} = 0.27,\Omega_{\Lambda,0} = 0.73$
and $H_0 \approx 72$ kms$^{-1}$Mpc$^{-1}$ if she leaves now. Note that
for this model,~\cite{Heyl05}, calculates a different value for the
proper time taken to traverse the same comoving distance and has his
rocketeer accelerating for $\approx 50$ years of proper time instead
to cover 99\% of the maximum distance with the same acceleration and
cosmology.

In Figure~\ref{fig:rocketeer}, we have calculated how far the
rocketeer reaches after travelling for various proper times at an
acceleration of $\epsilon = g = 9.81$ms$^{-2}$. As in~\cite{Heyl05},
we find that the majority of the maximum distance is covered between
$\approx$ 20--23 years, except when $\Omega_\Lambda \approx 0$. Thus,
for short trips, the greatest distance reached is largely unaffected
by the amount of dark energy in the universe. Up to $\approx$ 23
years, the total distance travelled is well described by a straight
line with a very small gradient over the range of energy densities,
which implies that the behaviour of the scale factor, for all
spatially flat FLRW models with a cosmological constant, is mostly
irrelevant for the distances covered by the rocketeer. It's tempting
to think that it is possible to just approximate the full solution
with special relativity up to a significant distance but the
discrepancy is actually quite large in comparison to the numerical
result (see Appendix for a detailed discussion).

\begin{figure}
\includegraphics[scale = 0.43, angle = 270]{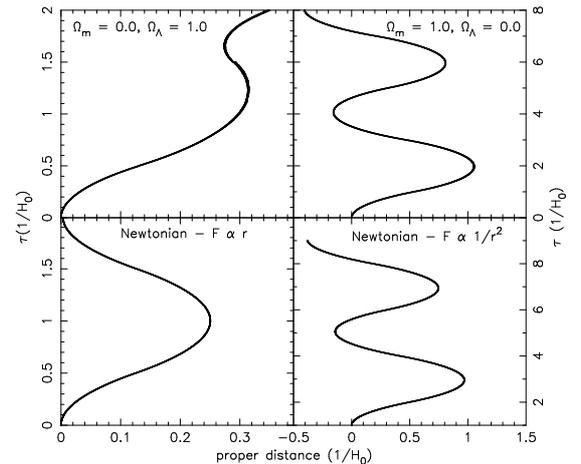}
\caption{Return paths of rocketeers in various cosmological models. A
single return journey consists of three stages: starting at $\tau =
0$, the rocketeer accelerates outwards at $\epsilon = $H$_0$ until
$\tau =$ H$_0^{-1}$, decelerates at $\epsilon = - $H$_0$ until $\tau =
3$H$_0^{-1}$, and finally accelerates at $\epsilon= $H$_0$ until $\tau
=4$H$_0^{-1}$ to attempt to return to the origin and stop at rest
relative to the cosmological fluid. All radial distances are measured
in proper distances.\label{fig:squiggly}}
\end{figure}

\begin{figure}
\includegraphics[scale = 0.33, angle = 270]{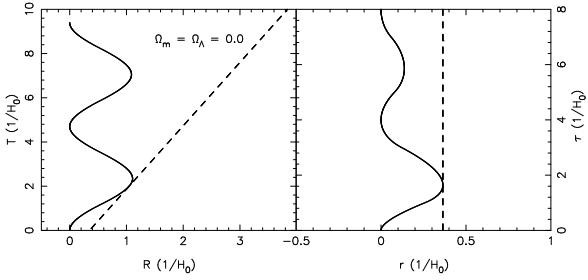}
\caption{Return paths in the Milne model or empty universe. Her journey is
structured in the same way as in Figure~\ref{fig:squiggly}. The above
plots were produced by repeating a return trip twice and we have also
included a comoving observer in dashed lines for comparison. The
rocketeer meets the comoving observer only once during her journey and
hence the path of a comoving observer in fully comoving coordinates
$(R,T)$ is tilted. The left panel shows her journey in conformal
coordinates, $R,T$, while the path on the right is in terms of
comoving coordinates. \label{fig:milne}}
\end{figure}

But suppose that a rocketeer wished to travel to a distant galaxy with
the intention of returning home; how would she have to fire her rocket
to bring her to her destination and back again in an expanding
universe? A single return journey as shown on the left panels in
Figure~\ref{fig:squiggly} is modelled by applying an acceleration that
is a piecewise function of proper time, such that:
\begin{equation}
\epsilon = \left\{
\begin{array}{rc}
1   &  {\rm ( 0 \leq \tau < 1/H_0) \:\: or \:\: (3/H_0 \leq \tau < 4/H_0)}, \\
-1  &  {\rm ( 1/H_0 \leq \tau < 3/H_0 )},
\end{array} \right. 
\label{return-acc}
\end{equation}
and this is repeated to produce the two return paths on the right. We
have used fairly long travel times and correspondingly very small
accelerations ($\epsilon = H_0 \approx 6.85 \times 10^{-11}g$) to
exaggerate any effects that the evolution of the scale factor may have
on the path of the rocketeer, since for large accelerations, the
trajectory asymptotes to a straight line soon after blast off. Three
universes have been chosen for their interesting properties;
Figure~\ref{fig:squiggly} shows the path of a rocketeer in a de Sitter
model ($\Omega_{m,0} = 0, \Omega_{\Lambda,0} = 1$) and an Einstein-de
Sitter model ($\Omega_{m,0} = 1, \Omega_{\Lambda,0} = 0$) in the top
two panels. We have also shown the path of the rocketeer under
Newtonian gravity (see~\cite{Whiting,Tipler_newt}) in the other two
panels.  The bottom right panel of Figure~\ref{fig:squiggly} uses the
familiar $1/r^2$ force law, but Newtonian gravity also admits a lesser
known linear force law for which Gauss' law still
holds~\citep{calder}. The inclusion of this term renders the Newtonian
force law into a form directly comparable to the Friedmann solution
with a cosmological constant in the limit of weak gravitational fields
as the coefficient of the term in $r$ behaves like $\Lambda$. Thus, we
can mimic a de Sitter model using an acceleration $ \propto r$ and an
Einstein de Sitter universe with an acceleration $ \propto 1/r^2$, the
results of which are analysed along the bottom row of
Figure~\ref{fig:squiggly}. Although, it is difficult to discern in the
left hand panels, each path in a GR-Newtonian pair has the same
frequency with respect to proper time. Additionally, we analyse the
motion of a rocketeer in a Milne or empty universe ($\Omega_{m,0} =
\Omega_{\Lambda,0} = 0$) under the same type of return journey in
Figure~\ref{fig:milne} for comparison, since in this scenario, the
spacetime is clearly static, as discussed below.



The top row of Figure~\ref{fig:squiggly} shows that in a FLRW
universes, the rocketeer does not return to the origin on her way back
home if her accelerations are symmetric with respect to her proper
time. We could choose to view this as an effect of `expanding space';
space grows bigger as she travels and so the rocketeer is being
carried away as if on a rubber sheet as the universe expands. Where
the analogy fails is for the matter dominated case - we would expect
that as the balloon is blown up, albeit at a decreasing rate, the
distance travelled is greater on the outbound leg than on the way
back. The crucial difference, of course, is the absence of gravity
which causes the rocketeer to overshoot in a matter dominated universe
and undershoot in a de Sitter universe. For motion restricted to the
radial direction, the relationship for the change in rapidity [$\chi =
\tanh^{-1}(u^r/u^t)$] with proper time is given by:
\begin{equation}
\frac{d\chi}{d\tau} = \epsilon - \frac{\dot{a}(\eta)}{a^2(\eta)}\sinh \chi,
\label{rapidity}
\end{equation}
(see ~\cite{Heyl05} for further details).~\cite{Heyl05} describes the
second term in Equation~\ref{rapidity} as a ``friction term" that
results from the expansion of the universe. In fact, $H(\eta) =
\frac{\dot{a}(\eta)}{a^2(\eta)}$, so we can view the significance of
this term in the above equation as a Hubble drag term that limits the
increase in the rapidity of the rocket.

The Hubble parameter describes the expansion history of the universe
and is given by:
\begin{equation}
H = H_0\sqrt{\Omega_{m,0} a^{-3} + \Omega_{\Lambda,0}}. 
\end{equation}
for a spatially flat FLRW models. We can immediately see that this is
going to be greater for accelerating universes than decelerating ones
at the same point in cosmic time. For instance, the Hubble parameter
of an universe with a mixture of matter and dark energy asymptotes to
to the de Sitter case as $t\rightarrow \infty$, so the maximum amount
of `drag' is always obtained for the rocketeer in a dark energy
dominated universe. For an empty universe, the second term on the
right hand side of Equation~\ref{rapidity} is zero. Thus the rocketeer
in a matter dominated universe where $\dot{a}(\eta)/a^2(\eta)$ is
decreasing, reaches a higher rapidity in the same proper time as a
rocketeer in the concordance model with accelerated expansion and
overshoots. However, the physics behind the analogy is simply that the
influence of changing energy densities on the path of the rocketeer
cannot be ignored~\citep{Lewis08}. Recently, there has been much
confusion in the literature regarding the physicality of expanding
space. It had been suggested by some~\citep{A07,A08, A09} that the
expansion of space is a physical phenomenon with observable
consequences rather a byproduct of a choice of coordinates. The above
asymmetry is apparently one of the reasons why \cite{A07,A08,A09}
support the notion of expanding space as a measurable effect - the
motion of particles in a FLRW universe are not purely described by
special relativity but are distorted by space time curvature. In fact,
the behaviour of a rocketeer in Newtonian gravity is very similar to a
rocketeer in a matter dominated universe. Previously, it has been
shown that, up to $\approx$ 100 Mpc, geodesic motion is qualitatively
well approximated by Newtonian gravity~\citep{Barnes06}. With an
accelerated observer, however, we cannot make a similar universal
statement to a give a scale on which Newtonian gravity is an accurate
approximation, because this is of course dependent on the rocketeer's
acceleration and the cosmology; this is evident from comparing the
left and right panels of Figure~\ref{fig:squiggly}. But as this figure
also demonstrates, both rocketeers overshoot; it is misleading to
attribute this behaviour in an Einstein-de Sitter model to the
stretching of space carrying away the rocketeer. It is sufficient to
say that the changing effect of gravity during the rocketeer's
journey, distorts her path such that the outward leg of her trip is
not the same as the inbound leg without having to invoke the analogy
of an expanding balloon.

In contrast, the behaviour of a rocketeer using a de Sitter model and
a linear Newtonian force in $r$ are only similar up to a proper time
of $\approx$ 0.8H$_0^{-1}$ after which the paths diverge. It is easier
to interpret this situation as an example of expanding space in
action, whereby the rockteer is being dragged away by the rapid
increase in the scale factor. But the rocketeer can only travel for as
long as the remaining conformal time in the de Sitter case, which can
not be mimicked in a Newtonian sense. The conformal factor $1/a(t)$
becomes negligible as $\eta \rightarrow$ H$_0^{-1}$ and at this time,
the Newtonian approximation breaks away from the GR picture. Again we
are presented with two choices for interpreting this situation; either
the exponential increase in the scale factor causes the rocketeer to
cover a greater proper distance with each trip or that $u^r$ increases
as the cosmic time approaches infinity to preserve the normalisation
of the four-velocity as $u^\eta \rightarrow 0$.  Each interpretation
is only separated by the choice of a metric, again we reiterate that
expanding space is a coordinate dependant effect as emphasised
in~\cite{Lewis08} and~\cite{bunn-2008}.  Furthermore, a careful
inspection of the Newtonian case reveals that the rocketeer really did
undershoot after all, although the amount by which this occurs is
still an order of magnitude away from its behaviour in GR.

Figure~\ref{fig:milne} corresponds to an universe empty of all
gravitating material in which the rocketeer is able to return to the
origin at rest in the same amount of proper time as the outward
journey. Although its scale factor is given by $a(t) = H_0 t$, this is
a static model developed from special relativity \cite{Milne} to
describe a universe consisting entirely of non-gravitating particles,
since it is possible to relate the FLRW metric of the empty universe
as a function of the comoving radial coordinate, $r$, and the cosmic
time $t$, to the metric of special relativity with coordinates $R, T$ via the
following transformation ~\cite{Rindlertext,Gron}:
\begin{equation}
R = t\sinh r, \qquad T = t\cosh r,
\label{empty_conformal}
\end{equation}
such that the line element becomes $dS^2 = -dT^2 + dR^2$ in two
dimensions. Thus we can consider the motion as being characterised by
special relativity; surfaces of constant $t$ have negative curvature,
but the curvature overall tends to zero as $t$ increases. Both axes of
the lower left panel of Figure~\ref{fig:milne} are plotted in fully
conformal coordinates, $R$ and $T$. Since this universe is actually
static, each oscillation with respect to $T$ retains the same shape.

\begin{figure}[t] 
\begin{center}
\includegraphics[scale = 0.3,angle=270]{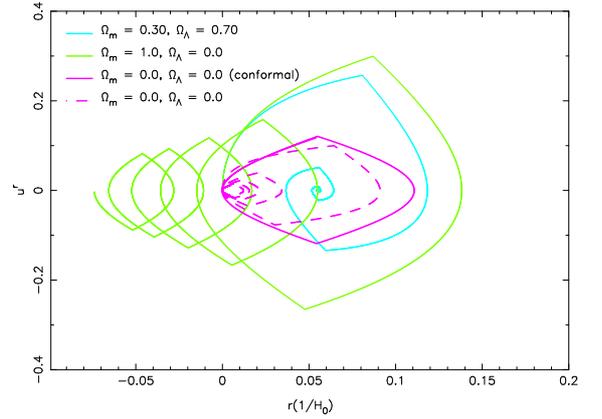}
\caption{Phase portraits for three different FLRW models. The paths
show five consecutive return journeys under the same conditions as in
Figure~\ref{fig:squiggly} and~\ref{fig:milne} for three cosmologies,
except the duration of each trip has been halved. Each universe has a
characteristic phase profile according to its long term behaviour. The
Milne model has been plotted in both synchronous and conformal
coordinates; as well as being scaled down to 10\% (solid) and 25\%
(dashed) of its original amplitude.
\label{fig:phase}} 
\end{center}
\end{figure} 

The long term behaviour of the rocketeer may be neatly encapsulated in
the phase portraits shown in Figure~\ref{fig:phase}. Again, we have
chosen three universes with very different cosmic evolutions: the
accelerated expansion of the concordance model (cyan), the eternally
coasting Einstein-de Sitter universe (green) and the SR metric in the
guise of a Milne universe (purple). These paths were produced using
the same acceleration scheme as Figures~\ref{fig:squiggly}
and~\ref{fig:milne} using Equation~\ref{return-acc} but have been
extended for five oscillations to provide a more detailed curve. Since
the acceleration of a rocketeer is symmetric about the origin for a
Milne universe (see Figure~\ref{fig:milne}), her phase portrait is a
conservative, closed orbit and she always returns back to the
origin. The phase portraits describing the other models are much more
informative: the path in the concordance cosmology clearly converges
towards a particular point as the rocketeer's journey is bounded by a
finite conformal time, while the phase portrait of a rocketeer in an
Einstein-de Sitter universe spirals to infinity and indeed would do so
if it had not been truncated to five oscillations. Thus the type of
orbit cleanly delineates the type of cosmological horizon present in
each model. However the point that the rocketeer tends towards in the
concordance model is strictly not an attractor, since if we started
from a different comoving coordinate, the destination of the rocketeer
would also be shifted, but an attractor would exist at $(\eta, u^r) =
(1.12339, 0)$ for the concordance model.

\begin{figure*}
\begin{center}
\includegraphics[angle=270,scale=0.3]{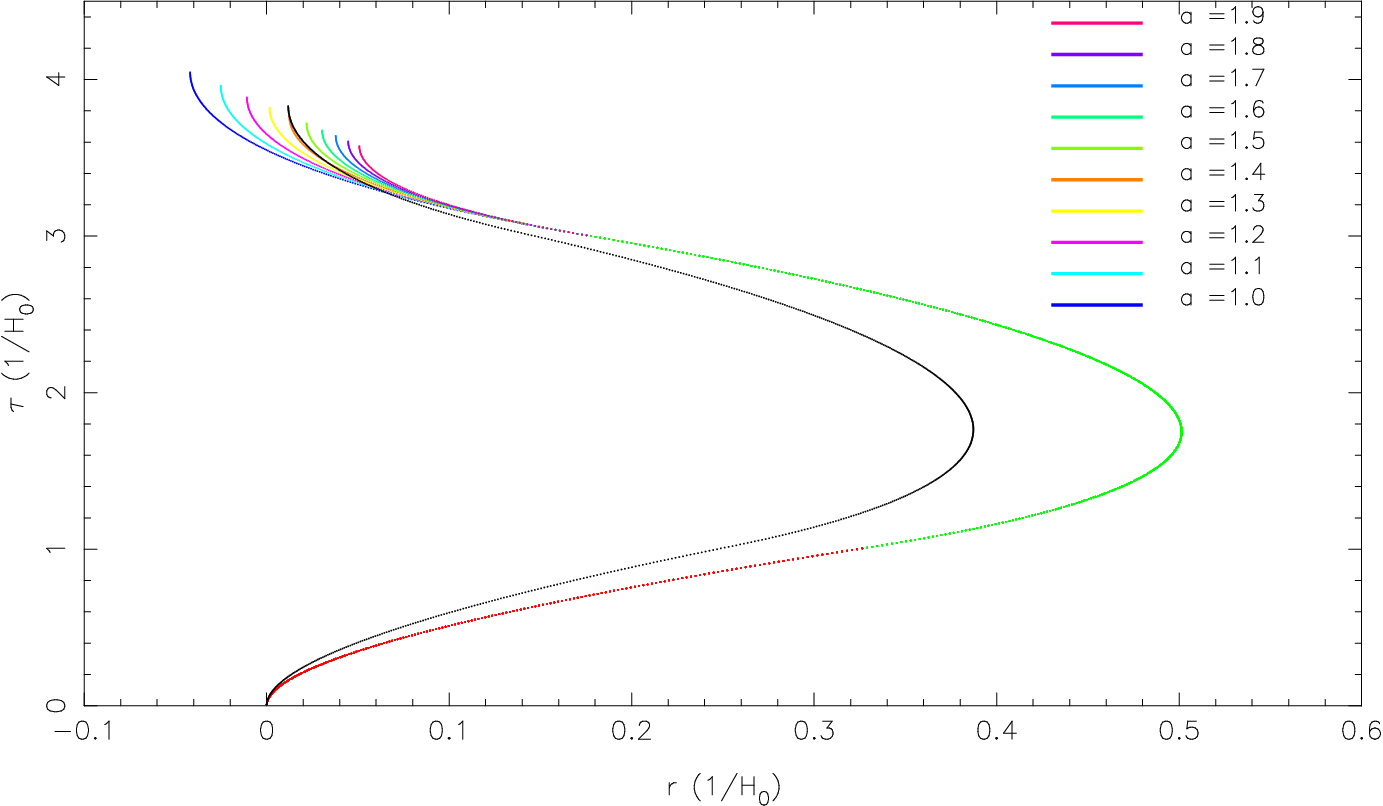}
\includegraphics[angle=270,scale=0.3]{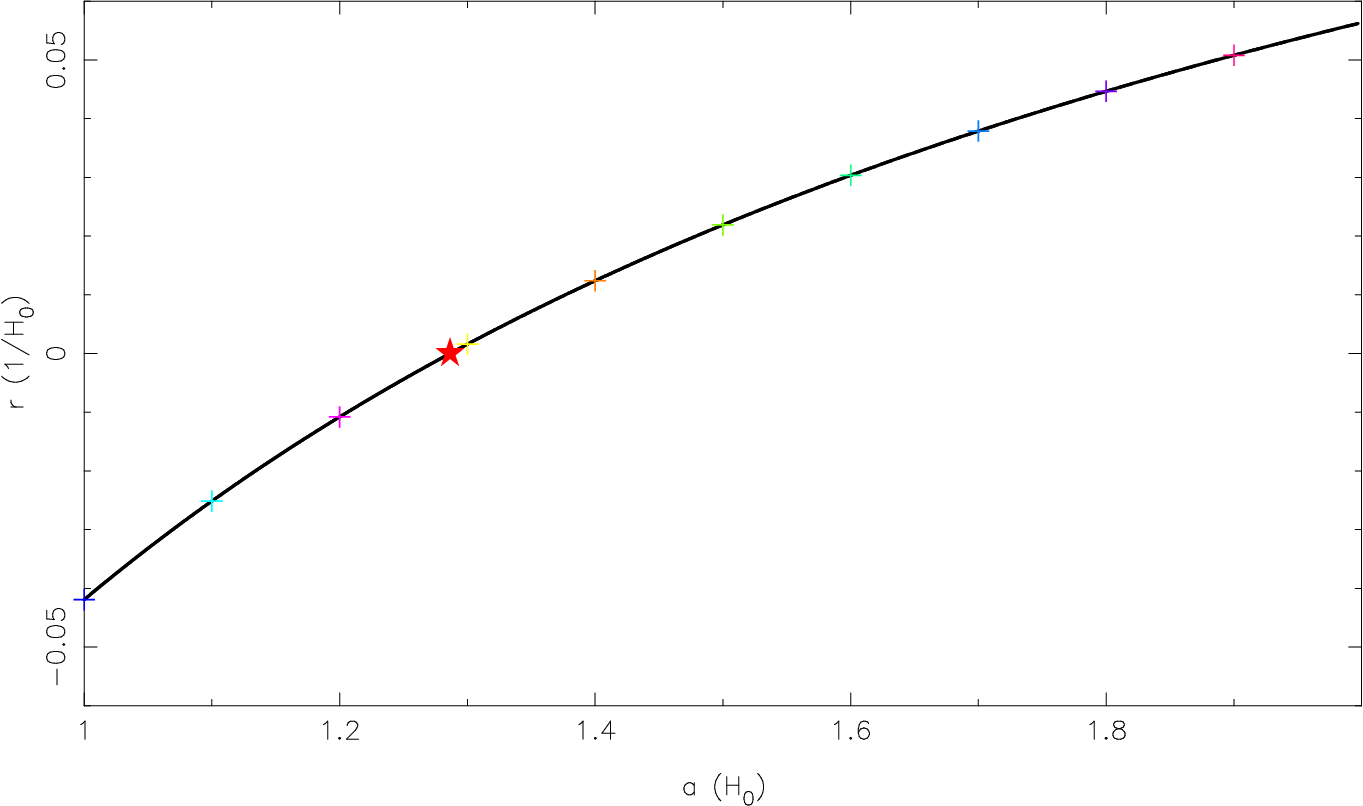}
\caption{Return paths for a rocketeer who accelerates outwards in the
same manner as in Figures~\ref{fig:squiggly} -- \ref{fig:phase} but
adjusts both the magnitude and length of the her acceleration during
the final stage of her journey such that $u^r = 0$ at the end. We have
used an Einstein-de Sitter model, which lacks the added complication
of a finite conformal time. The most favourable of these
paths is marked with a red star in the figure on the right, which shows the final
comoving radial coordinate of the rocketeer at $u^r = 0$ according to
her acceleration on her final leg; it appears that
the rocketeer is able to return to the origin if she accelerates at
$\epsilon =$ 1.421337 H$_0$ for 0.750610 H$_0^{-1}$. The black line
corresponds to a rocketeer who uses these parameters but in an
universe with the cosmological parameters $\Omega_{m,0} = 0.999$,
$\Omega_{\Lambda,0} = 0.001$. Coloured crosses on the right hand panel
correspond to paths on the left with the same colour.\label{fig:return}}
\end{center}
\end{figure*}

The dependence on cosmology in the equations of motion introduces
additional complexity that makes it unclear if a general solution that
can always return the rocketeer to the origin exists. If she is
allowed to change only the magnitude of her acceleration on the return
leg, the equations are over constrained and it is not possible to
return to the origin for any value of $\eta$. However adjusting both
its magnitude and duration rectifies this. Figure~\ref{fig:return}
shows an example of such a scenario in an Einstein-de Sitter universe
in which the rocketeer trials several values of $\eta$ and stops when
she reaches $u^r = 0$. The outward journey is remains the same as in
Figure~\ref{fig:squiggly}, the red curve corresponds to $\epsilon = 1$
for $H_0^{-1}$ and the green curve corresponds to $\epsilon = -1$ for
2$H_0^{-1}$, but on her final leg we have explored the effect of
varying the magnitude of $a$ as well as the length of proper time for
which she accelerates. The rocketeer would be able to return to rest
at the origin if she accelerated at $\epsilon =$ 1.28639 H$_0$ for
0.829959 H$_0^{-1}$. But her success relies heavily on being well
informed about the cosmology \emph{before} she commences her journey,
since any observation that she performs would be distorted by
relativistic aberration. The black curve in Figure~\ref{fig:return}
shows an unfortunate traveller who misjudges his cosmology by
0.1\%. Such a mistake would result in arriving $\approx$ 1.18490
$\times 10^{-2}$ H$_0^{-1} \: \approx $ 49.3 Mpc from Earth.

\section{Expanding Space: A Traveller's Guide}\label{coasting}
\begin{figure*}
\begin{center}
\includegraphics[scale = 0.3,angle=270]{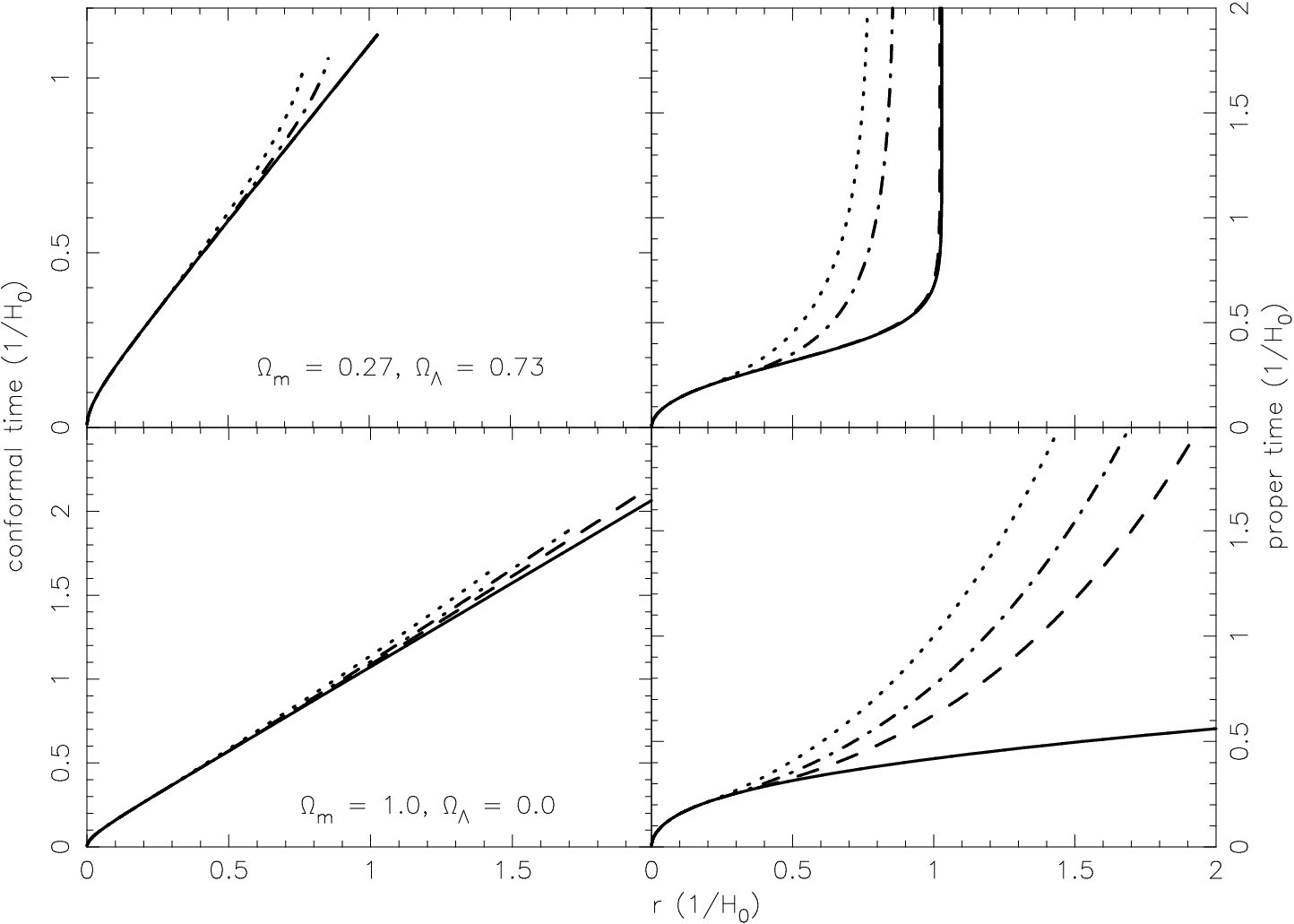}
\includegraphics[scale = 0.32,angle=270]{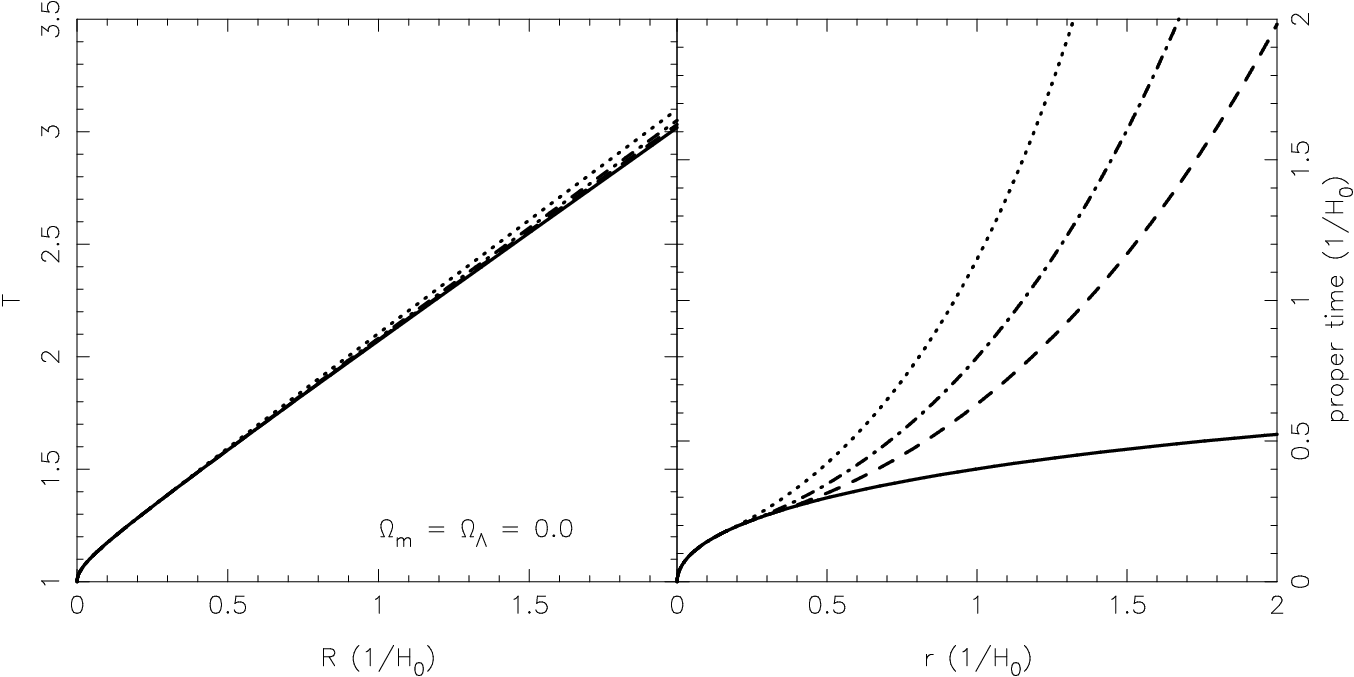}
\caption{Rocketeer vs.\ coaster for different universes, with their
cosmological parameters shown on the bottom right. We have plotted the
paths of several rocketeers with a constant outward acceleration of
$\epsilon=10$ H$_0$ who then switch off their engines when $u^r =2$
(dotted line), $u^r = 2.5$ (dot-dashed line) and $u^r = 3$ (dashed
line).  The solid line gives the path for a rocketeer who accelerates
outwards at $\epsilon = 10$ H$_0$ for the entire journey. A comparable
amount of total proper time passes for the rocketeers in each
universe.  As in Figure~\ref{fig:squiggly}, worldlines in the empty
universe have been plotted in fully conformal coordinates using
Equation \ref{empty_conformal}. Note that the axis has been truncated
at $R = 10$ H$_0^{-1}$ and $T = 10$ H$_0^{-1}$ for clarity.  Notice
that the coaster in dashed lines ($u^r_{max} = 3$) separates at a
later time in the concordance model than in any other universe because
it takes longer to reach the same velocity with cosmic acceleration.
\label{fig:coaster}} 
\end{center}
\end{figure*}

How can we best exploit the expansion of the universe such that future
space travellers can explore as much of the universe as they can in a
reasonable amount of proper time? ~\cite{Heyl05} considered this
question for a rocketeer with a constant acceleration, but achieving
this for the lifespan of a rocketeer may prove to be rather
unrealistic~\citep{Rindler60}. It is appropriate then, to consider the
amount by which the path of a rocketeer diverges from that of a
`coaster', a traveller of intermittent acceleration, if at all.
Na\"{i}vely we might expect that the Hubble drag term would force the
motion of the coaster back into the general expansion of the universe,
as discussed in~\cite{Barnes06} for geodesic motion.  In
Figure~\ref{fig:coaster}, we have plotted the results from solving for
such a situation in three different universes with the same parameters
as those in Section~\ref{squiggly}. The path of rocketeer travelling
with $\epsilon=10$ H$_0$ is shown as a solid line with a set of
rocketeers who accelerate at $\epsilon=10$ H$_0$ initially, but then
switch off their engines at $u^r = 2$ (dotted line), $u^r = 2.5$
(dot-dashed line) and $u^r = 3$ (dashed line). The difference in the
final comoving coordinates achieved by the coasters and rocketeer is
not immediately apparent from Figure~\ref{fig:coaster}, since the
paths of the rocketeer have been truncated for the middle and bottom
panels to maintain a reasonable scale between the three worldlines.

It is apparent from the conformal diagrams on the left of
Figure~\ref{fig:coaster}, that while the coaster and the rocketeer
have similar paths in all the conformal representations, after the
coasters turns off their engines, they experience a significantly
smaller amount of time dilation when compared to the rocketeer. For a
particular comoving radial coordinate in the diagrams on the right,
the proper times measured by each vary dramatically and only the
coasters in the concordance model are a comparable distance from the
rocketeer in comoving coordinates (but are separated by a significant
proper distance). While the coasters may keep up with the rocketeer in
terms of their comoving spatial separation, the times they would
measure would be considerably greater than that on the clock carried
by the rocketeer. However, if we take slices at a constant conformal
time, both the rocketeer and the coasters in all three universes are
at similar radial coordinates, which implies that the coasters skim
along the light cone approached by the rocketeer.

With a more realistic acceleration of $\epsilon=g$, a traveller in the
concordance model, who starts to coast after reaching a three-velocity
of $0.8c$ at a comoving distance of 0.189 pc, can expect to travel
$\approx 41.1$ pc after a total proper time of 100 years has
elapsed. Even with a maximum three-velocity of $0.99c$, the coaster
can only accelerate up to 1.67 pc after a year, and must drift for a
further 99 years to reach $\approx 208.5$ pc.  This is much less than
the furthest distance travelled by a rocketeer in the same proper time
($\approx$ 4.7 Gpc) and coasting appears to be of little use in
intergalactic exploration. It seems that in this scenario, in which
the distance travelled by the coaster before switching off her engines
is much less than the maximum conformal time, it is not practical for
a rocketeer to alter between accelerated and geodesic motion. However,
if a spacecraft of limited acceleration is the only means of travel
available, a coaster could maximise their final comoving distance by
accelerating up to $\approx 26$ years (or $\approx 23$ years for a
lower bound). There is a preferred duration for which the rocketeer
should accelerate, to maximise the distance that she covers while
minimising the amount of proper time spent travelling. Recall from
Figure~\ref{fig:rocketeer} that after this period, the rocketeer adds
little to her journey, at least for $\Lambda$-dominated
universes. This can be understood from a comparison of the results
shown in Figure~\ref{fig:rocketeer} with the panels on the right in
Figure~\ref{fig:coaster}; if the coaster switches to non-accelerated
motion too early in her journey, she is not experiencing the rapid
jump in the comoving distance travelled by the rocketeer occurring
between $ 20 \lesssim \tau \lesssim 26$ years (in
Figure~\ref{fig:rocketeer}) and $0.3 \lesssim \tau \lesssim 0.4
H_0^{-1}$ (in Figure~\ref{fig:coaster}). However, this is only
applicable to universes containing with a finite conformal time; a
coaster in an Einstein-de Sitter universe will always be much more
disadvantaged than a rocketeer.

Recently,~\cite{Krauss} have concluded that as the
behaviour of our Universe becomes increasingly dominated by dark
energy, future astronomers will infer the existence of a static
universe as described by the de Sitter metric, written its original
form:
\begin{equation}
ds^2 = (1-r^2/R^2)dt^2-\frac{dr^2}{1-r^2/R^2}-r^2d\Omega^2,
\end{equation}
where $R$ is the radius of a 4D hypersphere in 5D Euclidean
space~\citep{deSitter}. While a coordinate transformation links this
metric to the synchronous gauge, that it can be written in terms of a
cosmological constant will be nothing more than a mathematical
curiosity as cosmic acceleration is unlikely to be supported by
observational evidence in the very distant future. All objects not
gravitationally bound to our Galaxy will have receded such that
neither the Hubble expansion or the elemental abundances in quasar
spectra will be discernible and the intensity of the CMB will have
been redshifted below our observational capacity. While an observer
stationed on Earth will not detect the presence of dark energy, a
rocketeer travelling beyond the Local Group could provide a simple
means of determining the correct cosmological parameters. In
Section~\ref{squiggly}, it was shown that with constant accelerated
motion, cosmological distances can be travelled well within a human
lifespan. Although the conformal time bounds how far the
rocketeer can travel such that she may not be able to reach a
sufficiently distant destination to observe any interesting structure,
the most useful experiment that she could carry out would be to travel
in a return path with a constant acceleration or act as a coaster. As
demonstrated in the preceding section, the path of rocketeer is
extremely sensitive to the amount of dark energy contained in the
universe and in fact, there is quite a significant deviation between
two paths with cosmologies that differ by only 0.1\% (see previous
section). However, while this experiment will only take a few years
for the rocketeer, cosmologists on Earth will have to wait a very long
time for their results.



\section{Conclusions}\label{conclusions}
Accelerated motion provides a means for exploring a significant
fraction of our universe within the remaining conformal time in a
reasonably short time. In this work, we have examined solutions of the
equation of motion corresponding to accelerated round-trips in a
spatially flat Universe. It has been demonstrated that these paths
through spacetime are marked by the action of cosmic evolution, and
the manner in which cosmological parameters constrain the properties
of the acceleration for the purpose of completing the return trip has
been investigated. Moreover, accelerated paths illustrate the
pedagogical features of conformal diagrams: this is evident in the
characterisation of curves in the phase portraits and also in the
failure of the rocketeer to return exactly the origin. Using these
results, we have discussed the capacity for accelerated paths to cover
large distances across the Universe. By comparing different rates of
acceleration, we have demonstrated that particular choices of motion
can be chosen to optimise particular scientific outcomes. That is,
while coasting through a FLRW universe by using an intermittent
acceleration is an ineffective way of exploring our Universe,
accelerated motion may provide the only means of determining a correct
description.

\section*{Acknowledgements}
The authors wish to acknowledge support from ARC Discover Project
DP0665574 and Matt Francis for his comments and suggestions.

\section*{Appendix: Calculations}
We can solve for the first order perturbative solution to the full GR
equation of motion by using the method of multiple scales
[c.f.~\cite{Bender}]. After performing a conformal transformation and
introducing the standard relationships between the four-velocity and
four-acceleration in Equations~\ref{frw_norm_expanded} -
\ref{frw_perp_expanded}, we obtain the following differential equation
for the conformal time as a function of the proper time of the
rocketeer:
\begin{equation}
\frac{du^\eta}{d\tau} = -\frac{da(\eta)}{d\eta}\frac{1}{a(\eta)}{u^\eta}^2 + \frac{da(\eta)}{d\eta}\frac{1}{{a(\eta)}^3} + \frac{\epsilon}{a(\eta)}\sqrt{{a(\eta)}^2{u^\eta}^2 - 1}.
\label{eqn:cont}
\end{equation}
Of course the radial coordinate of the rocketeer is simply given by
solving the corresponding geodesic equation in $r$ as follows:
\begin{equation}
\frac{du^r}{d\tau} = -2\frac{da(\eta)}{d\eta}\frac{1}{a(\eta)}u^ru^\eta + \epsilon u^\eta 
\label{eqn:conr}
\end{equation}
To solve these, we introduced a perturbation in $\epsilon$ and a new
timescale such that
\begin{eqnarray}
\eta = \eta_0(\tau,\bar{t}) + \epsilon\eta_1(\tau,\bar{t}) + \epsilon^2\eta_2(\tau,\bar{t}) + O(\epsilon^3), \\
r = r_0(\tau,\bar{t}) + \epsilon r_1(\tau,\bar{t})+ \epsilon^2r_2(\tau,\bar{t}) + O(\epsilon^3),
\end{eqnarray}
\noindent where $\bar{t} = \epsilon \tau$, such that $\eta$ and $r$
are not truly functions of two variables. We can then write down
Equations~\ref{eqn:cont} and \ref{eqn:conr} as a system of PDEs. The
first order solution for $u^\eta$ is given by:
\begin{equation}
u^\eta = \cosh(\epsilon\tau/2)/a(\eta)
\end{equation}
Substituting this into Equation~\ref{eqn:conr} then gives an integral
solution for $u^r$:
\begin{equation}
u^r = \epsilon/a^2(\eta)\int a(\eta)\cosh(\epsilon\tau/2) \: d\tau 
\end{equation}
For certain forms of the scale factor, we can explicitly integrate
these equations again for the trajectory of the rocketeer in terms of
the proper time that she measures. If we choose a power-law solution
for the scale factor for simplicity, $a(t) =
{(t/t_0)}^\frac{2}{3(1+w)}$, where $w$ is the equation of state
parameter, and use Equation~\ref{eqn:conformal} to relate $\eta$ to $t$,
we can write down $a(\eta)$ as follows:
\begin{equation}
a(\eta) = \frac{1+3w}{3t_0(1+w)}\left[\eta +  \frac{1+3w}{3t_0(1+w)}\right]^{\frac{2}{1+3w}},
\end{equation}
for $a(\eta)|_{\eta = 0}= 1$. The solutions for $\eta_0$ and $r_0$
obtained from integrating $u^\eta$ and $u^r$ are:
\begin{equation}
\eta_0 = \frac{3(1+w)}{1+3w}\left(t_0^{\frac{2}{3(1+w)}}\left[2\sinh(\epsilon\tau/2)+t_0\right]^{\frac{1+3w}{3(1+w)}} - t_0 \right), \\
\label{eqn:soln_powerlaw_eta}
\end{equation}
\begin{multline}
r_0 = at_0\frac{3(1+w)}{5+3w}\int^\tau_0 ( f(\bar{\tau})^\frac{1+3w}{3(1+w)} \\
    -f(\bar{\tau})^\frac{-4}{3(1+w)} ) \: d\bar{\tau} \:,
\label{eqn:soln_powerlaw_r}
\end{multline}
where $f(\tau)= \frac{2}{\epsilon t_0}\sinh(\epsilon\tau/2)+1$. Note
that it is not correct to directly substitute $\eta_0$ into the
equation for the normalisation of the 4-velocity to derive an
expression for $u_r$, because the first-order solutions do not
necessarily satisfy these conditions exactly, so the result obtained
is not strictly a first order perturbation.

\begin{figure}[t] 
\begin{center}
\includegraphics[scale = 0.3,angle=270]{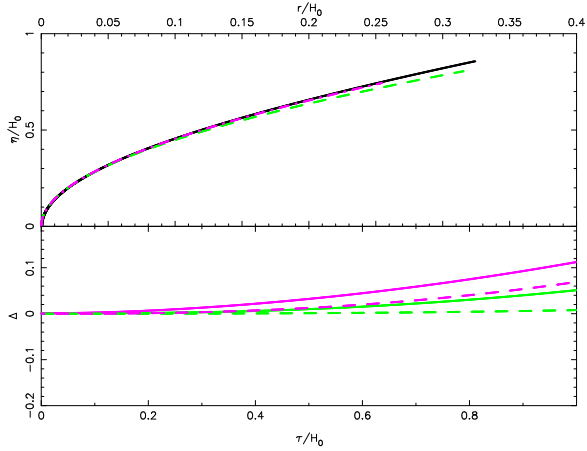}
\caption{A comparison between various solutions for accelerated paths
through an Einstein de Sitter model ($\Omega_{m,0} = 1,
\Omega_{\Lambda,0} = 0, k = 0)$, with $\epsilon = 1$. In the
top panel, we have shown the paths that are obtained from numerically
integrating Equation~\ref{ageo} (solid black), the first order solution of this
equation in conformal coordinates
(Equations~\ref{eqn:soln_powerlaw_eta} and~\ref{eqn:soln_powerlaw_r},
green dashed) and the solution in special relativity
(Equation~\ref{eqn:milne}, pink dot-dashed). The bottom panel shows
the error, $\Delta$, between the numerical result and the solution
from perturbation theory and the error between the numerical result
and the solution in special relativity for both the conformal time
(solid lines) and the radial distance (dashed lines) of the
rocketeer. The colours have been chosen to match those of the top
panel.
\label{fig:rocket_comp}} 
\end{center}
\end{figure}

It is important to remember that these are first order solutions and
hence at most accurate to $\tau = O(1/\epsilon)$. The accuracy of
these solutions is demonstrated in Figure~\ref{fig:rocket_comp}, in
which we compare them to numerical results and the solution in special
relativity. The difference between the solutions contained in
Figure~\ref{fig:rocket_comp} and the numerical result is smaller
than the same comparison for special relativity. Thus, if we were
looking for a way to approximate the full behaviour in general
relativity, the use of perturbative methods on the geodesic equation
is preferable to using special relativity for even moderately distant
destinations, in this instance, anything further than 0.4 H$_0^{-1}$.

\begin{thebibliography}{}


\expandafter\ifx\csname natexlab\endcsname\relax\def\natexlab#1{#1}\fi
\expandafter\ifx\csname bibnamefont\endcsname\relax
  \def\bibnamefont#1{#1}\fi
\expandafter\ifx\csname bibfnamefont\endcsname\relax
  \def\bibfnamefont#1{#1}\fi
\expandafter\ifx\csname citenamefont\endcsname\relax
  \def\citenamefont#1{#1}\fi
\expandafter\ifx\csname url\endcsname\relax
  \def\url#1{\texttt{#1}}\fi
\expandafter\ifx\csname urlprefix\endcsname\relax\def\urlprefix{URL }\fi
\providecommand{\bibinfo}[2]{#2}
\providecommand{\eprint}[2][]{\url{#2}}

\bibitem[{\citenamefont{Abramowicz et~al.}(2007)}]{A07}
\bibinfo{author}{\bibnamefont{Abramowicz}, \bibfnamefont{M.~A.}},
  \bibinfo{author}{\bibnamefont{Bajtlik}, \bibfnamefont{S.}},
  \bibinfo{author}{\bibnamefont{Lasota}, \bibfnamefont{J.-P.}},
  \bibnamefont{and}
  \bibinfo{author}{\bibnamefont{Moudens}, \bibfnamefont{A.}},
  \bibinfo{journal}{Acta Astronomica} \textbf{\bibinfo{volume}{51}},
  \bibinfo{pages}{139} (\bibinfo{year}{2007}).

\bibitem[{\citenamefont{Abramowicz}(2008)}]{A08}
\bibinfo{author}{\bibnamefont{Abramowicz}, \bibfnamefont{M.~A.}},
  \bibinfo{journal}{New Astronomy Reviews} \textbf{\bibinfo{volume}{51}},
  \bibinfo{pages}{799} (\bibinfo{year}{2008}).

\bibitem[{\citenamefont{Abramowicz et~al.}(2008)}]{A09}
\bibinfo{author}{\bibnamefont{Abramowicz}, \bibfnamefont{M.~A.}},
  \bibinfo{author}{\bibnamefont{Bajtlik}, \bibfnamefont{S.}},
  \bibinfo{author}{\bibnamefont{Lasota}, \bibfnamefont{J.-P.}},
  \bibnamefont{and}
  \bibinfo{author}{\bibnamefont{Moudens}, \bibfnamefont{A.}},
  \bibinfo{journal}{ArXiv e-prints}  (\bibinfo{year}{2008}),
  \eprint{0812.3266}.

\bibitem[{\citenamefont{Barnes et~al.}(2006)}]{Barnes06}
\bibinfo{author}{\bibnamefont{Barnes}, \bibfnamefont{L.~A.}},
  \bibinfo{author}{\bibnamefont{Francis}, \bibfnamefont{M.~J.}},
  \bibinfo{author}{\bibnamefont{James}, \bibfnamefont{J.~B.}}, \bibnamefont{and}
  \bibinfo{author}{\bibfnamefont{G.~F.} \bibnamefont{Lewis}},
  \bibinfo{journal}{MNRAS} \textbf{\bibinfo{volume}{373}}, \bibinfo{pages}{382}
  (\bibinfo{year}{2006}).

\bibitem[{\citenamefont{Bender and Orszag}(1978)}]{Bender}
\bibinfo{author}{\bibnamefont{Bender}, \bibfnamefont{C.~M.}},
  \bibnamefont{and} \bibinfo{author}{\bibnamefont{Orszag}, \bibfnamefont{S.~A.}
  }, \emph{\bibinfo{title}{Advanced {M}athematical {M}ethods for {S}cientists and {E}ngineers: {A}symptotic {M}ethods and {P}erturbation {T}heory"} (\bibinfo{publisher}{McGraw-Hill}
  Press}, \bibinfo{address}{New York}, \bibinfo{year}{1978}).

\bibitem[{\citenamefont{Bunn and Hogg}(2008)}]{bunn-2008}
\bibinfo{author}{\bibnamefont{Bunn}, \bibfnamefont{E.~F.}}, \bibnamefont{and}
  \bibinfo{author}{\bibnamefont{Hogg}, \bibfnamefont{D.~W.}},
  \emph{\bibinfo{title}{The kinematic origin of the cosmological redshift}}
  (\bibinfo{year}{2008}), \urlprefix\url{arXiv.org:0808.1081}.

\bibitem[{\citenamefont{{Calder} and {Lahav}}(2008)}]{calder}
\bibinfo{author}{\bibnamefont{Calder}, \bibfnamefont{L.}} \bibnamefont{and}
  \bibinfo{author}{\bibnamefont{Lahav}, \bibfnamefont{O.}},
  \bibinfo{journal}{Astronomy and Geophysics} \textbf{\bibinfo{volume}{49}},
  \bibinfo{pages}{010000} (\bibinfo{year}{2008}), \eprint{0712.2196}.





\bibitem[{\citenamefont{{de Sitter}}(1917)}]{deSitter}
\bibinfo{author}{\bibnamefont{de Sitter}, \bibfnamefont{W.}},
  \bibinfo{journal}{Koninklijke Nederlandse Akademie van Weteschappen
  Proceedings Series B Physical Sciences} \textbf{\bibinfo{volume}{19}},
  \bibinfo{pages}{1217} (\bibinfo{year}{1917}).


\bibitem[{\citenamefont{Francis et~al.}(2007)}]{Francis}
\bibinfo{author} {\bibnamefont{Francis}, \bibfnamefont{M.~J.}},
  \bibinfo{author}{\bibnamefont{Barnes}, \bibfnamefont{L.~A.}},
  \bibinfo{author}{\bibnamefont{James}, \bibfnamefont{J.~B.}}, \bibnamefont{and}
  \bibinfo{author}{\bibnamefont{Lewis}, \bibfnamefont{G.~F.}},
  \bibinfo{journal}{PASA} \textbf{\bibinfo{volume}{24}}, \bibinfo{pages}{95}
  (\bibinfo{year}{2007}).

\bibitem[{\citenamefont{Gr{\o}n}(2006)}]{Gron}
\bibinfo{author}{\bibnamefont{Gr{\o}n}, \bibfnamefont{{\O}.}},
  \bibinfo{journal}{Eur. J. Phys.} \textbf{\bibinfo{volume}{27}},
  \bibinfo{pages}{561} (\bibinfo{year}{2006}).

\bibitem[{\citenamefont{Gr{\o}n and Elgar{\o}y}(2007)}]{GE}
\bibinfo{author}{\bibnamefont{Gr{\o}n}, \bibfnamefont{{\O}.}}, \bibnamefont{and}
  \bibinfo{author}{\bibnamefont{Elgar{\o}y}, \bibfnamefont{{\O}.}},
  \bibinfo{journal}{Am. J. Phys.} \textbf{\bibinfo{volume}{75}},
  \bibinfo{pages}{151} (\bibinfo{year}{2007}).

\bibitem[{\citenamefont{Heyl}(2005)}]{Heyl05}
\bibinfo{author}{\bibnamefont{Heyl}, \bibfnamefont{J.~S.}},
  \bibinfo{journal}{Phys. Rev. D} \textbf{\bibinfo{volume}{72}},
  \bibinfo{pages}{107302} (\bibinfo{year}{2005}).

\bibitem[{\citenamefont{Hobson et~al.}(2006)}]{Hobson}
\bibinfo{author}{\bibnamefont{Hobson}, \bibfnamefont{M.~P.}},
  \bibinfo{author}{\bibnamefont{Efstathiou}, \bibfnamefont{G.}},
  \bibnamefont{and} \bibinfo{author}{\bibnamefont{Lasenby}, \bibfnamefont{A.~N.}
  }, \emph{\bibinfo{title}{General {R}elativity{:} {A}n
  {I}ntroduction for {P}hysicists}} (\bibinfo{publisher}{Cambridge University
  Press}, \bibinfo{address}{Cambridge}, \bibinfo{year}{2006}).

\bibitem[{\citenamefont{Infeld and Schild}(1945)}]{Infeld}
\bibinfo{author}{\bibnamefont{Infeld}, \bibfnamefont{L.}} \bibnamefont{and}
  \bibinfo{author}{\bibnamefont{Schild}, \bibfnamefont{A.}},
  \bibinfo{journal}{Phys. Rev.} \textbf{\bibinfo{volume}{68}},
  \bibinfo{pages}{250} (\bibinfo{year}{1945}).

\bibitem[{\citenamefont{Krauss and Scherrer}(2007)}]{Krauss}
\bibinfo{author}{\bibnamefont{Krauss}, \bibfnamefont{L.~M.}} \bibnamefont{and}
  \bibinfo{author}{\bibnamefont{Scherrer}, \bibfnamefont{R.~J.}},
  \bibinfo{journal}{Gen. Relativ. Gravit.} \textbf{\bibinfo{volume}{39}},
  \bibinfo{pages}{1545} (\bibinfo{year}{2007}).

\bibitem[{\citenamefont{Lewis et~al.}(2008)}]{Lewis08}
\bibinfo{author}{\bibnamefont{Lewis}, \bibfnamefont{G.~F.}},
  \bibinfo{author}{\bibnamefont{Francis}, \bibfnamefont{M.~J.}},
  \bibinfo{author}{\bibnamefont{Barnes}, \bibfnamefont{L.~A.}},
  \bibinfo{author}{\bibnamefont{Kwan}, \bibfnamefont{J.}}, \bibnamefont{and}
  \bibinfo{author}{\bibnamefont{James}, \bibfnamefont{J.~B.}},
  \bibinfo{journal}{MNRAS} \textbf{\bibinfo{volume}{388}}, \bibinfo{pages}{960}
  (\bibinfo{year}{2008}).

\bibitem[{\citenamefont{Lewis et~al.}(2007)}]{Lewis07}
\bibinfo{author}{\bibnamefont{Lewis}, \bibfnamefont{G.~F.}},
  \bibinfo{author}{\bibnamefont{Francis}, \bibfnamefont{M.~J.}},
  \bibinfo{author}{\bibnamefont{Barnes}, \bibfnamefont{L.~A.}},
  \bibnamefont{and} \bibinfo{author}{\bibnamefont{James}, \bibfnamefont{J.~B.}},
  \bibinfo{journal}{MNRAS} \textbf{\bibinfo{volume}{381}}, \bibinfo{pages}{L50}
  (\bibinfo{year}{2007}).

\bibitem[{\citenamefont{{Loeb}}(2002)}]{Loeb}
\bibinfo{author}{\bibnamefont{Loeb}, \bibfnamefont{A.}},
  \bibinfo{journal}{Phys. Rev. D} \textbf{\bibinfo{volume}{65}},
  \bibinfo{pages}{047301-1} (\bibinfo{year}{2002}).

\bibitem[{\citenamefont{Milne}(1932)}]{Milne}
\bibinfo{author}{\bibnamefont{Milne}, \bibfnamefont{E.~A.}},
  \bibinfo{journal}{Nature} \textbf{\bibinfo{volume}{130}}, \bibinfo{pages}{9}
  (\bibinfo{year}{1932}).


\bibitem[{\citenamefont{{Perlmutter et al.}}(1999)}]{Perlmutter}
\bibinfo{author}{\bibnamefont{Perlmutter}, \bibfnamefont{S.}},
\bibinfo{author}{\bibnamefont{Aldering}, \bibfnamefont{G.}},
\bibinfo{author}{\bibnamefont{Goldhaber}, \bibfnamefont{G.}},
\bibinfo{author}{\bibnamefont{Knop}, \bibfnamefont{R.~A.}},
\bibinfo{author}{\bibnamefont{Nugent}, \bibfnamefont{P.}},
\bibinfo{author}{\bibnamefont{Castro}, \bibfnamefont{P.~G.}},
\bibinfo{author}{\bibnamefont{Deustua}, \bibfnamefont{S.}},
\bibinfo{author}{\bibnamefont{Fabbro}, \bibfnamefont{S.}},
  \bibnamefont{et~al.}\bibinfo{journal}{ApJ} \textbf{\bibinfo{volume}{517}},
  \bibinfo{pages}{565S} (\bibinfo{year}{1999}).

\bibitem[{\citenamefont{{Riess et al.}}(1998)}]{Riess}
\bibinfo{author}{\bibnamefont{Riess}, \bibfnamefont{A.~G.}},
\bibinfo{author}{\bibnamefont{Filippenko}, \bibfnamefont{A.~V.}},
\bibinfo{author}{\bibnamefont{Challis}, \bibfnamefont{P.}},
\bibinfo{author}{\bibnamefont{Clocchiatti}, \bibfnamefont{A.}},
\bibinfo{author}{\bibnamefont{Diercks}, \bibfnamefont{A.}},
\bibinfo{author}{\bibnamefont{Garnavich}, \bibfnamefont{P.~M.}},
\bibinfo{author}{\bibnamefont{Gilliand}, \bibfnamefont{R.~L.}},
\bibinfo{author}{\bibnamefont{Hogan}, \bibfnamefont{C.~J.}},
  \bibnamefont{et~al.}\bibinfo{journal}{ApJ} \textbf{\bibinfo{volume}{116}},
  \bibinfo{pages}{1009} (\bibinfo{year}{1998}).

\bibitem[{\citenamefont{Rindler}(1960)}]{Rindler60}
\bibinfo{author}{\bibnamefont{Rindler}, \bibfnamefont{W.}},
  \bibinfo{journal}{Phys. Rev.} \textbf{\bibinfo{volume}{119}},
  \bibinfo{pages}{2082} (\bibinfo{year}{1960}).

\bibitem[{\citenamefont{Rindler}(2006)}]{Rindlertext}
\bibinfo{author}{\bibnamefont{Rindler}, \bibfnamefont{W.}},
  \emph{\bibinfo{title}{{Relativity: Special, General, Cosmological, 2nd Ed.}}}
  (\bibinfo{publisher}{Oxford {U}niversity {P}ress}, \bibinfo{address}{Oxford},
  \bibinfo{year}{2006}).


\bibitem[{\citenamefont{{Tipler}}(1996)}]{Tipler_newt}
\bibinfo{author}{\bibnamefont{Tipler}, \bibfnamefont{F.~J.}},
  \bibinfo{journal}{MNRAS} \textbf{\bibinfo{volume}{282}}, \bibinfo{pages}{206}
  (\bibinfo{year}{1996}).

\bibitem[{\citenamefont{{Whiting}}(2004)}]{Whiting}
\bibinfo{author}{\bibnamefont{Whiting}, \bibfnamefont{A.~B.}},
  \bibinfo{journal}{The Observatory} \textbf{\bibinfo{volume}{124}},
  \bibinfo{pages}{174} (\bibinfo{year}{2004}), \eprint{arXiv:astro-ph/0404095}.



\end{thebibliography}
\end{document}